\documentstyle[12pt,epsf]{article}
\begin{document}
\baselineskip 22pt plus 2pt
\begin{center}
{\bf On the Biological Advantage of Chirality}\footnote{to be
published in ``Advances in Biochirality'', Eds. G. Palyi,
C. Zucchi \& L. Caglioti, Elsevier Science S.A.}
\\ \ \\
by \\ \ \\
G. Gilat\\
Department of Physics, Technion\\
Haifa 32000, Israel \\
e-mail: \ gilat@physics.technion.ac.il
\\ \ \\ \ \\
\end{center}

{\bf Abstract}

The presence of chirality in the main molecules of life may well be not just
a structural artifact, but of pure biological advantage.  The possibility of
the existence of a phenomenon of a special mode of interaction, labeled
as "chiral interaction"
(CI), for which structural chirality is a necessary condition, is the
main reason for
such an advantage.  In order to  demonstrate such a possibility, macroscopic chiral
devices are introduced and presented as analogies for such an interaction.
For this purpose it is important to make a clear distinction between geometric and
physical chiralities, where the latter are capable to perform chiral interactions
with various media.  Apart from chirality, a few other structural
elements are
required. In particular, the presence of an interface that separates between
the chiral device and the medium with which it is interacting. The physical chirality
is build into this very interface where chiral interaction is taking place.  On
a molecular level, soluble proteins in particular, the active medium is the presence
of random ionic motion in the aqueous solvent.  As a result of chiral interaction a
certain perturbation, or current, is generated and flowing along the coils of $\alpha$-helix
structure in one preferred direction out of two possible ones.  A model for such a
chiral interaction is presented and a few significant consequences are pointed
out.  In particular, it is important to emphasize the time-irreversible feature
of chiral interaction, which leads to its non-ergodic nature that is to be
considered a necessary condition for evolutionary processes in biomolecules.
As yet there exists no direct experimental evidence for the validity of chiral interaction
on a molecular level, but there are quite a few indirect supporting evidences.
In particular, there exists an experimental result by Careri et al. who found a
clear linkage between the free protonic motion in the hydration layer of proteins
and their enzymatic activity.  A few direct experiments for verifying the validity of
chiral interaction on molecular level are proposed hereby for hydrophobic amino
acids at the water-air interface, where chiral interaction may take place.
Among these there is also a new approach of applying a SQUID to detect a weak
magnetic field that is associated with the chiral interaction effect.  If proven
right, chiral interaction may open new approaches and possibilities for better
understanding of the rather complex autocatalytic function of soluble proteins.
\pagebreak

\noindent{\bf 1. Introduction}

The phenomenon of structural chirality has been recognized since the early 19th
century when Arago [1] and Biot [2] demonstrated the effect of optical activity in
quartz crystals.  It was Louis P{asteur [3] who first observed chirality on a
molecular level and he referred to this effect as dissymmetry.  It was Kelvin [4]
who first addressed this phenomenon as "chirality", since a chiral structure
of an object does not necessarily imply a total lack of symmetry of such an object.
According to Kelvin a given object or set is chiral if and only if it cannot be made
to superimpose, or overlap, exactly its mirror image by any continuous transformation.
(i.e. by any rotation and/or translation).  The concept of chirality is of
geometric nature in principle, but currently this very term is also being used in
other domains such as high energy physics, which may cause semantic problems in its
usage.  In addition to this it is also important to address chirality within
its own dimensionality.  The space we live in is 3-dimensional (3d) so that chirality
is usually regarded as a 3d property.  In fact, such property exists also in
2-dimensional (2d) space, that is, within a plane.  This can be demonstrated by two
asymmetric triangles, one being the mirror image of another.  Although these
two triangles are geometrically identical, they cannot be made to superimpose
one another by any rotation and/or translation within the plane they are contained in.
In order to make them coincide with one another, it is necessary to take one
of these triangles out of this plane, rotate it around in 3d space
and then bring it back
into  the same plane.  Then it becomes possible to make them overlap precisely with
one another. Such a triangle is an example of 2d chiral figure or object.  Any 2d chiral
object is considered to be achiral in 3d space.  As a matter of fact, such a
consideration is not limited to 2d space.  In principle, also a 3d chiral object
"can be made" to coincide with its mirror image by "taking it out into a 4-dimensional
(4d) space, rotating it around there and then bringing it back to 3d space".  Such an
operation is mathematically or imaginatively possible, but not practically.
the reason for this is that our physical space is of 3-dimensions only.

In recent years there has been a considerable progress in the development and understanding
of the mathematical aspects of chirality [5-6].  Among other things, there exist several
attempts [7-9] of quantifying this very concept. These attempts are still having
certain problems of uniqueness,  so that a question such as: "What is the
most chiral triangle?" obtains at least three different answers [10,11].
As a matter of
fact, most approaches of treating chirality from mathematical aspects employ the
geometric view-point of chirality, which hardly contains any physical meaning.
In reality, the concept of chirality has a significant physical basis and this is
meant to be one of the main features of the present article.  For this reason
it becomes necessary to distinguish between so-called geometric and physical
chiralities [11,12].  Geometric chirality is referred to the shape of pure geometric
bodies or sets, such as triangles or tetrahedrons or any arbitrary shapes in
2- or 3-dimensional spaces.  Physical chirality is referred to the shape of any
distribution in space of a physical property such as mass, charge, energy potential, or even
quantum properties such as electronic wave-function distribution throughout a
molecule, or any other quantum mechanical property.  If such a distribution does not
contain any reflection symmetry plane, then, by analogy, it is to be referred to
as a chiral physical distribution.  There exist two major distinctions
between physical
and geometric objects.  Geometric bodies are distributed homogeneously
in space,
whereas physical properties may be of varying densities, such as
mass distribution,
which is not necessarily homogeneous in space.  This difference may
well complicate
the mathematical aspects of physical chiralities.  A second distinction
of major
physical significance is the possibility of interaction that may exist between
a distribution of physical property and some medium that happens
to be present at
the same vicinity.  In the case of molecular structures it may be necessary to
distinguish between various different physical distributions such as nuclear
masses, (i.e. atomic locations within the molecule), or electronic distributions that
are presented by the wave-function distributions.  The possibility of a
chiral distribution of a physical property is the  origin of the biological
advantage of chirality.  In order to clarify this statement it may be
helpful to look first at macroscopic objects where chirality plays a necessary
and useful role in their function.

\noindent{\bf 2.  Macroscopic Chiral Devices}

An intriguing question that is being repeated in may scientific publications concerns
the `left' (L) and `right' (D) identity of chiral molecules, such as: "Why
are amino-acids L and not D?". There exist several speculations that try to
answer this interesting question.  A more constructive question to be asked presently is:
"why are the molecules of life chiral?", and this is regardless of their being L
or D.  Is there any biological advantage in their chiral nature when compared to
achiral molecules?  And the answer given here to this question is: "Yes".
The source of such an advantage comes from a specific type of interaction for which
chirality is a necessary condition. Such an interaction is to be labeled as:
"chiral interaction" (CI) and it has already been treated in various publications
[11,13-17].

Chiral interaction (CI) is not typical of molecular structure only, and there
exist various macrodevices which function according to the same principle.
The simplest and most spectacular one is the windmill.  When wind blows at the rotors
of a mill, it ` knows' immediately in which direction to rotate, clockwise or
anti-clockwise.  If the windmill, in particular its vanes, were symmetric, it would
not be able to `make up its mind' to select a definite rotational direction.
The shape of vanes of a windmill, where contact is made with the wind, is
designed to break the L-D symmetry, i.e., it is chiral.  This structure is of
useful dynamical advantage which enables the mill to transform energy from the
wind to, say, a rotating millstone.  Another simple mechanical device is a
rotating water sprinkler, where the medium interacting with the sprinkler is the
flow of water.  The next example is somewhat more sophisticated because
it depends on a
different mode of chirality.  This device is a variant of the
Crookes' radiometer
(see figure 1).  The active medium in this case is light radiation.  The
element of chirality here is in the difference of the colors of the rotating blades,
being black and silver, respectively, and this is a special example of
physical chirality.  When light is shining on silver faces it is
reflected away, whereas
it is absorbed by the black ones.
\pagebreak
\begin{figure}[htb]
\centerline{\epsfxsize=10cm \epsffile[130 220 480 560]{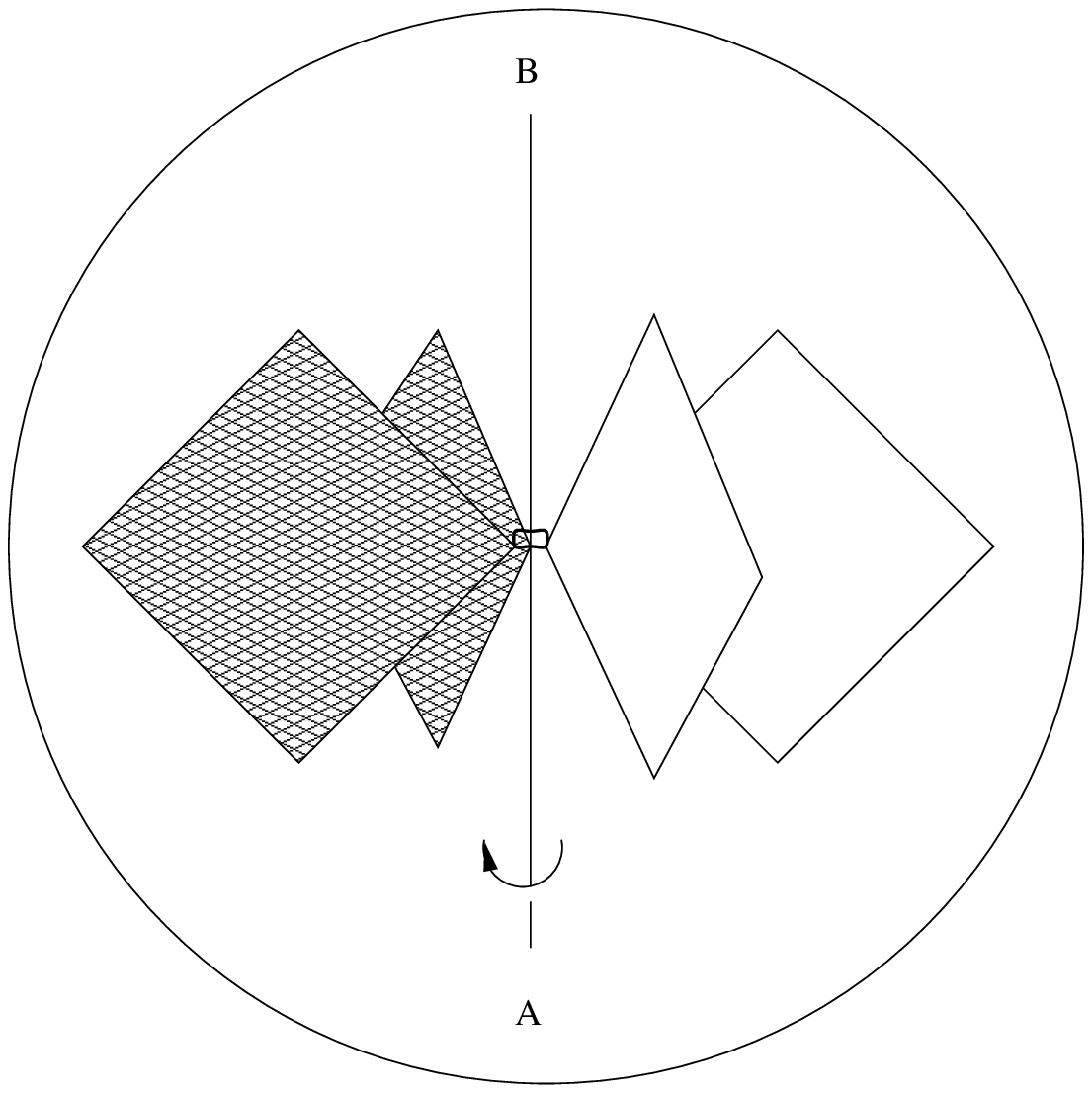}}
\caption{A variant of Crookes' radiometer offers an
example of chiral interaction. The asymmetry in the optical absorption
coefficient between the black and the silver vanes generates a temperature
difference between these vanes when light is shining at the device. The air
close to the black vanes expands as a result, which pushes the vane around
the axis AB in the preferred direction of the black vane. The physical
chirality is built into this device by the distribution of the black and
silver colors, i.e. by the difference in the optical absorption coefficients
in the vanes.}
\end{figure}

\pagebreak

  This causes the black faces to become warmer
than the silver ones, which in turn expands the air in their close vicinity,
so that the black wing is pushed backwards.  The selection of the sense of rotation
of this type of Crooke's radiometer is made by the variance of the colors of the
vanes, that is, by their respective optical absorption coefficient.  The
physical chirality in this device is presented by the distribution
of the optical absorption
coefficient, not by the shape of the vanes which is achiral.

All the examples presented here are of mechanical nature, i.e., the
effect of chiral interaction (CI) results in a mechanical rotation in one
preferred direction out of two possible ones around a given axis of the
device. This is so because the source of the interaction, i.e. the medium,
usually is external to the chiral device. In the case of an electric device
which generates a static current flowing in one preferred direction out of
two possible ones, the source of the interaction may be embedded within the
device. This is the case, for instance, of an electric cell which consists
of two different electrodes coming in contact with an electrolyte. It is
obvious that in order to reverse the direction of the current it is necessary
to interchange the two electrodes with one another, but this does not
necessarily require any chiral operation. This is so because the source
of the current flow is internal, so that the structure of the device can
be completely symmetric, as is the case of a cylindric battery. In the case
of an electric thermocouple, the operation can still be regarded as CI since
the source of the interaction, i.e. the temperature difference, is external
to the device.


Before proceeding to molecular systems, it becomes useful to summarize the main
features of CI in macrodevices. In all these examples there exists a specific active
medium with which each chiral device is interacting, and this happens always at an
interface separating the device from the medium, where physical chirality is
built into this interface.  CI is a process by which energy is transferred from the
medium to the chiral device, where one out of two possible modes of rotational motion
is selected and generated. Such a feature of selectivity can be considered as
a mode of organization.  Two modes of different natures, namely,
mechanical or electric
motions, can be created within the device.  The electric mode, i.e.
the current,
is of practically massless charge motion flowing along a conductor.

\noindent{\bf 3. \ Chiral Interaction in Molecular Systems}\\
{\bf 3a. General Features and Physical Background.}

The origin of CI in molecular systems is in the motion of free ions in
aqueous solutions.  This is to be considered as the necessary medium for CI,
whereas the interacting `chiral device', i.e. the chiral molecule,
must include
certain specific features which enable it to function as a `chiral device'.
These conditions fit well soluble proteins that contain electric dipole moments
that interact electrostatically with moving ions in the solvent.  Owing to the
large dielectric constant of water ($\epsilon = 81$), the range of CI is limited
to approximately $(10-20)\AA$.  The permanent electric dipole moment
[18] of the
protein molecule consists of the amino (NH$^+$) and the carbonyl
(CO$^-$) groups along the
peptide bonding chains.  This dipole moment is presented schematically by PN (see figure 2).  Let us now examine the motion of a

\pagebreak
\begin{figure}[htb]
\centerline{\epsfxsize=7cm \epsffile[94 205 422 556]{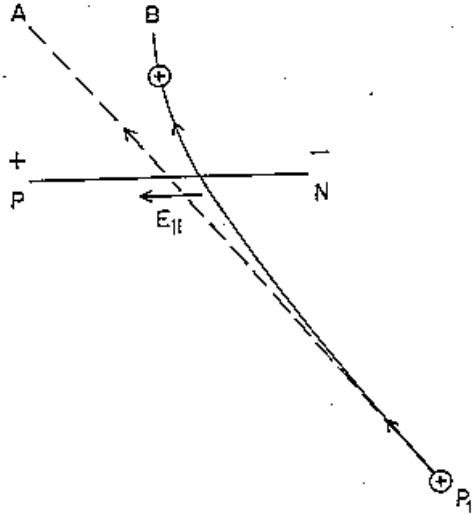}}
\caption{The deflection of a positive ion $P_i$
from a linear track, presented by $P_1A$, into a curved track $P_1B$ due
to an electric dipole moment PN.  The curved motion of $P_1$ induces a
non-zero time-averaged electric field $E_{\parallel}$ along PN.  A negative
ion (not shown here)is deflected in the opposite direction, but, being of
opposite sign to $P_1$, it will induce $E_{\parallel}$ in the same direction
of $P_1$.}
\end{figure}
\pagebreak

 positive ion P$_1$ approaching
PN.  Let $E_{\parallel}(t)$ be the time dependent component of an electric field
$\vec{E}$ induced by P$_1$ along PN.  The time average of $E_\parallel(t)$ is
given by:
\begin{eqnarray}
E_\parallel = \frac{1}{T} \int^T_0 E_\parallel (t) dt   
\end{eqnarray}

In the absence of any interaction P$_1$ moves along a straight line P$_1$A so that
for $T\rightarrow \infty$, we have $E_\parallel \rightarrow 0$.  Since PN
is in fact an electric dipole moment, the actual track of the ion P$_1$
is now presented by the orbit of P$_1$B rather than P$_1$A.  This is so because
the ion is being deflected away from P and attracted by N.  For this
shape of an
orbit, not being on a straight line,
$E_\parallel \neq 0$ for $T\rightarrow \infty$,
and it is pointing in the direction of $P$.  Let us now look for the case of
a negative moving ion, not shown in the figure.  It is also deflected by
the electric dipole but in the opposite direction in comparison to P$_1$.
Being also of opposite charge, it will induce an electric field $E_\parallel$ in the
same direction as the positive ion.  In other words, $E_\parallel$
remains in the same direction for positive and negative ions alike.
Another fact that is of significance is the independence of the motional direction
of P$_1$.  For instance, if P$_1$ moves backward on the same track in the
opposite direction, it will still produce $E_\parallel$ in the same direction
as before.

The electric field $E_\parallel$, caused by the deflection of random ionic
motion in the solvent, due to the existence of a permanent dipole moment,
becomes a source of repetitive perturbation of the molecular system.
The system
responds to such an external perturbation, according to the Le Chatelier law,
in an attempt of maintaining an equilibrium state within the system,
i.e., it evokes
an internal perturbation along closed loops of chemical bonds in an
attempt to
neutralize the external perturbation. Such an internal perturbation, or `current',
selects one out of two possible directions of motion as in the case
in macrodevices,
in particular, the electric cell.  This effect is to be regarded as
`Chiral Interaction'
(CI), on a molecular level scale [19].  Since the molecules involved in such
an interaction, i.e. soluble proteins, are considerably heavier than the deflected
ions in the solvent, any mechanical effects of CI can be neglected in comparison
to its electric effect.

\noindent {\bf 3b. The Necessity of an Interface}

In the description of chiral macrodevices the necessary existence of an interface,
where CI is taking place, has been emphasized.  Surprisingly enough, such an
interface becomes also necessary for molecular chiral systems.  This is shown
schematically in Figure 3.  If moving ions approach a closed ring attached to an
electric dipole PN, from all possible directions, the induced perturbation, i.e. CI, averages
out to zero.  This is so because I$_A$ cancels out I$_B$ on the average, so that
the net induced perturbation becomes zero on the average.
For this reason, in order
to maintain a non-zero CI, it becomes necessary to limit the access of ions approaching
the ring to about half of the space surrounding it, i.e., limit the access of
P$_2$, in comparison to P$_1$.  This can be accomplished by an interface
(denoted by S),
that prevents the access of free ions, i.e. P$_2$ from the space above $S$, so that
CI can go on in one preferred direction, i.e. I$_A$, for an indefinite length of
time.

In nature there exist several interfaces.  Most obvious one is the solvent-air
interface, which is macroscopic.  At this interface there exists a strong

\newpage
\begin{figure}
\centerline{\epsfxsize=7cm \epsffile[94 280 420 555]{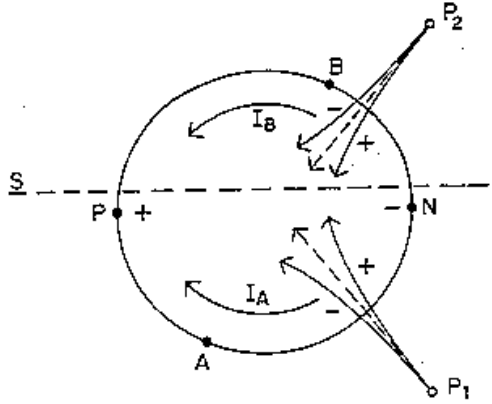}}
\caption{Schematic representation of why it is necessary
to have an interface in order to maintain a non-zero chiral interaction.
Let $P_1$ be a moving ion approaching an electric dipole moment PN fixed,
for simplicity, on the diameter of a ring of closed chemical bonds.
$P_1$ is deflected from its linear track according to its charge sign, and
induces an electric field $E_{\parallel}$ and a resultant chiral current
$I_A$ in the ring. A similar ion $P_2$ approaches from above and induces
$I_B$ in the opposite direction along the ring, so that on the average
there exists no CI. In order to maintain a non-zero CI it is necessary to
stop or limit the approach of ions from one of these two semicircles. This
can be accomplished by a separating interface represented by S.}
\end{figure}
$~~~$
\newpage

change, or gradient $\vec{\bigtriangledown} c$, in the ionic strength $c$ of the solvent.
This provides for a variation $|\vec{\bigtriangledown} c \cdot d|$ in the
ionic strength across a loop of chemical bonds such as
$C_\alpha \cdot COO^-\cdot NH^+_3$,
of a hydrophobic amino acid molecule [13] attached to the water
surface, where $d\approx \ {\rm few} \ \AA$.  The origin of the
electric dipole
moment along this loop is due to the existence of a zwitterion in
amino acids in
the  presence of water.  A sufficiently strong gradient in the ionic strength at
the close vicinity of the water-air interface can maintain CI for
molecules such
as amino acids that are aligned normal to this interface.  Another natural interface
of molecular size exists at the surface of globular proteins.
This interface
intervenes between  the solvent, where free ionic motion persists, and the
the interior part of the molecule where no ionic motion exists.  Owing to this
interface, CI can occur for globular proteins in the bulk of the solvent.
Somewhat  surprising and encouraging is the observation that soluble proteins must
become globular in order to function as enzymes [20].  This fact may indicate
a possible linkage between CI and enzymatic activity of proteins, a possibility to
be further discussed in what follows.  Looking at the secondary
$\alpha$-helix
structure of proteins, each segment of such a loop, containing the NH$^+ \dots$CO$^-$
bond, can be considered as a `chiral element' with which free ions in the solvent
are interacting (see figure 4). \ This mode of interaction, being CI on a molecular
level, is schematically shown and described in figure 5.

\noindent {\bf 3c. Formal Treatment of Chiral Interaction}

Next, a formal treatment of the intrinsic molecular perturbation, or `chiral
current' I$_c$ is presented for a single chiral element.

\newpage
\begin{figure}
\centerline{\epsfxsize=6cm \epsfbox{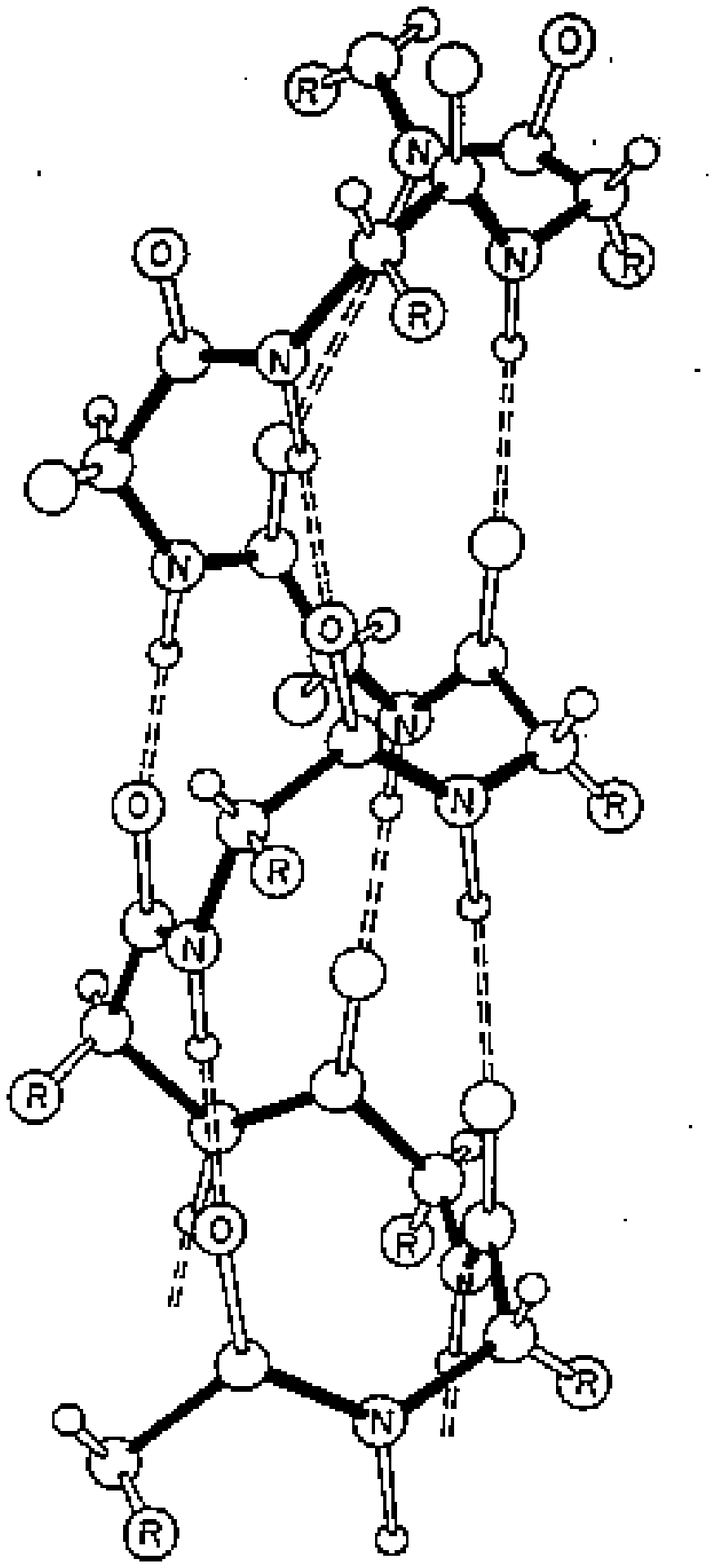}}
 \caption{A segment of an $\alpha$-helix structure
 which can be considered as a segment in a globular protein. The H-bond
 connecting two successive coils are presented by broken lines and they
 are all oriented in the same direction, i.e. from O(-) \ to \ N(+). The
 H-bonds together with the covalent bonds connecting N and O around the loop
 are considered as a `chiral element'. (Original drawing by Irwin Geis).}
\end{figure}

\newpage
\begin{figure}
\centerline{\epsfxsize=7cm \epsfbox{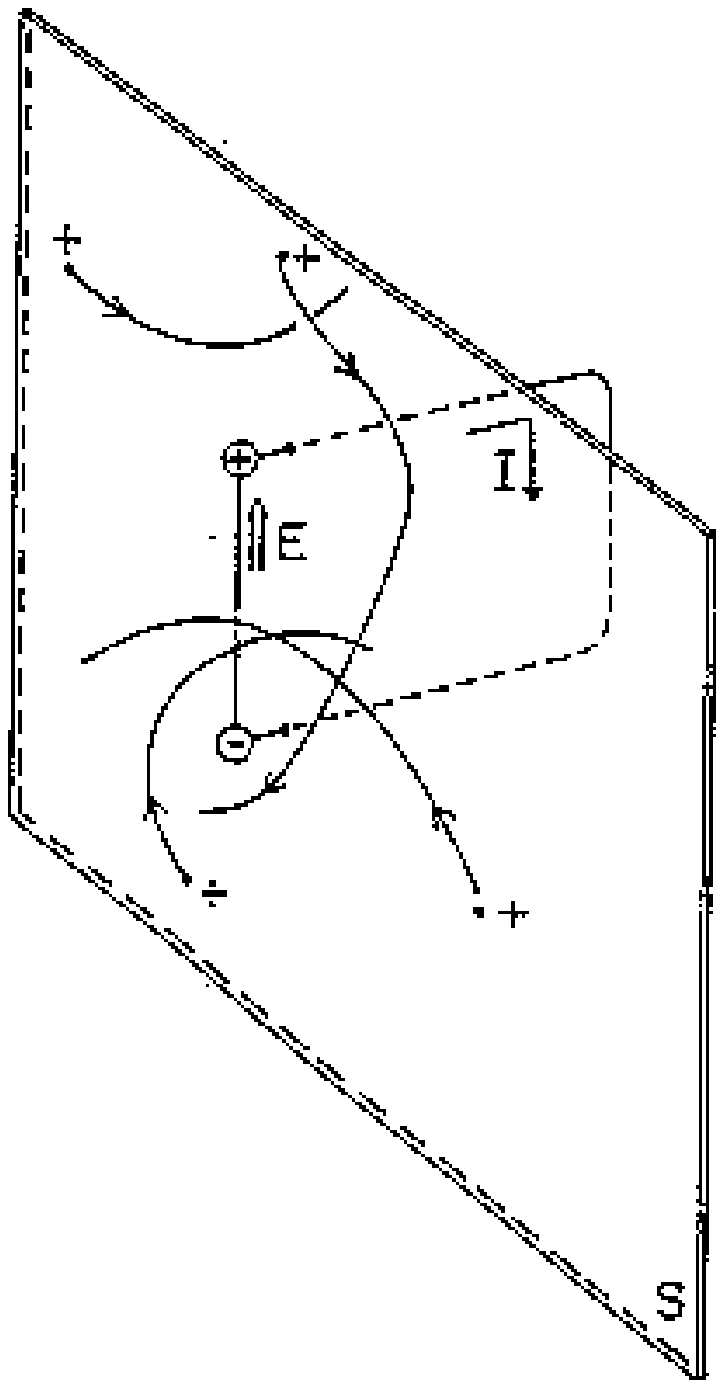}}
\caption{A schematic representation of a chiral
element in a globular protein. The electric dipole moment sticking into
the solvent, that is, on the left hand side of the interface S, represents
the H-bonds interconnecting the coils in the $\alpha$-helix. The possible
motion of ions (only positive ones are shown) in the solvent is shown, and
so is the induced electric field E that generates the chiral current I in
the chiral element. The interface S represents the surface of the globular
protein so that mobile ions exist only on the left side of S.}
\end{figure}
\newpage

This is carried out by
applying the Langevin equation [13,21]:
\begin{eqnarray}
\frac{d I_c}{dt} = - \gamma I_c + F(t) 
\end{eqnarray}
where $\gamma = \tau^{-1}_R$ is the attenuation or dissipation constant of
$I_c$ and $\tau_R$ is the relaxation time of the chiral perturbation or current
I$_c$. F(t) represents a series of stochastic events of random
approaches of moving ions
in the solvent to the chiral element presented in figure 4 or 5.  F(t) is presented
mathematically by:
\begin{eqnarray}
F(t) = \tau^{-1}\sum \alpha_i\delta(\theta - \theta_i)\exp[-\gamma(t-t_i)] 
\end{eqnarray}
where $\alpha_i$ is the i-th current, or perturbation increment occurring at
time $t_i$.  $\theta = t/\tau$, $\theta_i=t_i/\tau$ and $\tau$ is the stochastic
time constant, being the average time gap between two successive stochastic
events of solvent ions approaching the chiral element. The solution of
equation 2 yields:
\begin{eqnarray}
I_c = \sum \alpha_i\Theta(\theta-\theta_i)\exp[-\gamma(t-t_i)] 
\end{eqnarray}
where $\Theta(\chi)$ is a step-function.  The time-average $\langle I_c\rangle$
for long periods $T_i>>\tau$ and $(T_1-T_2)>>\tau$ is given by
\begin{eqnarray}
\langle I_c \rangle = \frac{1}{T_2-T_1}\int^{T_2}_{T_i} I_c(t)dt 
\end{eqnarray}
For a steady-state situation it is assumed that $\langle d I_c/dt\rangle \cong
0$, so that [11,17]:
\begin{eqnarray}
\langle I_c \rangle = \frac{\alpha}{\gamma\tau} = \frac{\alpha\tau_R}{\tau} 
\end{eqnarray}
where $\alpha = \langle \alpha_i\rangle$ is the average chiral current perturbation
generated per each stochastic event, and $\alpha$ is proportional to the magnitude of
the electric dipole-moment in the peptide bond.  $\tau$ is inversely proportional
to $c$, which represents the ionic strength of the solvent,
so that $\langle I_c\rangle$ is proportional to the product
$\alpha\tau_R$c, that is:
$\langle I_c\rangle \propto \alpha \tau_R c$.  From equation 6 it can
be deduced that if $\tau_R << \tau$ then
$\langle I_c \rangle \rightarrow 0$.  In other
words, if the relaxation time $\tau_R$ of the chiral current perturbation
is much shorter than the average stochastic time $\tau$, then CI vanishes.
The physical meaning of this is that the ionic strength $c$ must be sufficiently
large to maintain the chiral interaction CI. \\

\noindent{\bf 3d. Estimates and Conclusions}

In the absence of any experimental data concerning CI in molecular systems,
it is still not easy to make reliable estimates for the size of the effect of CI.
It is obvious that, owing to the large mass ratio of the protein relative to a
moving ionic mass, only CI of the second kind can be invoked, i.e. of electronic
or massless motion.  The estimate of the energy involved in this process of
CI is of the order of ($10^{-2}-10^{-3})$kT per stochastic event (17), which signifies
it is a `subthermal effect'.  The relatively small energy contents does not
mean that such an effect can be ignored.
On the contrary, the smallness of this effect, as well as the degree of its sophistication,
may provide for information transfer throughout a complex molecule
such as protein,
and may be significant for its specific enzymatic function.  This aspect will
be further discussed in what follows.  The necessary ingredients of CI
on a molecular level may be now listed as:
\begin{enumerate}
\item The existence of a chiral element (or molecule) that includes an electric
dipole moment as part of the chiral structure;
\item The presence of a polar solvent that is sufficiently rich in mobile ions;
\item The existence of closed loops of chemical bonds in the chiral element;
and
\item The presence of an interface that separates between two media that differ
largely in their respective ionic strength and/or in their ionic mobilities in the
close vicinity of a chiral element.
\end{enumerate}

It should be emphasized that CI  in biomolecules is still a hypothesis awaiting for experimental
verification.  If confirmed, it may become of considerable significance as a
mode of self-organization that is pertinent to the intrinsic control of enzymatic
activity of proteins.  It should also be stressed that all the ingredients
required for CI occur in nature for globular proteins as well as for amino acids
oriented at the water-air interface.  It is also of interest to point out the relevance, or
necessity, of the presence of nitrogen atoms in the structure of amino acids
and proteins.  The physical asymmetry between the amino NH$^+$ and the carbonyl
CO$^-$ groups provides for the existence of an electric dipole moment which is a
necessary condition for CI.  Hydrocarbons alone, lacking the presence of N, cannot
participate in the generation of CI.  it is important to notice that physical
effects of chirality have well been known for long [3], and they involve mainly
optical phenomena, such as optical activity [22] and related effects.
These effects are to be considered as `chiral scattering'  rather than
`chiral interactions' because they concern {\it only} the effect of chiral
systems on electromagnetic radiation.  In the case of CI, the main effect to be
considered is the generation of an intrinsic perturbation {\em within} the chiral
system, which is ignored in the case of chiral scattering.  In addition to this,
CI is not limited to optical effects only, as is the case in the phenomena of chiral
scattering.  Needless to add that the distinction between chiral scattering and CI
does not mean that there is no relation between these two concepts.  The opposite
is correct.\\

\noindent{\bf 4. General Features and the Advantages of Chiral Interaction}\\
{\bf 4a Thermodynamic and Space-Time Symmetry Considerations}

Thermodynamical aspects of CI have already been considered [11,16-17].
The basic
effect of CI, where a molecule is being perturbed and thus
is developing a persistent
current in a single selected direction, is thermodynamically uncommon.
Such a
perturbation can be considered as an excited state of indefinitely long lifetime.
This means that however small is the energy of such a perturbation, the system will
never completely reach thermal equilibrium.  That is, as long as the
chiral system is
surviving.  Such a phenomenon has certain peculiar aspects and in order to better
appreciate them let us now look back at the example of a windmill.  While so doing,
macroscopic effects such as friction and other energy dissipative effects are
ignored. What we see now is a significant space-time symmetry feature of
CI, being of
{\it time-irreversible} nature, in contrast to many common mechanical
examples.  For instance,
if flow of time is reversed, a satellite moving in space `will move' on the same track
in the opposite direction.  This is so because of the nature of the forces acting on
it in space.  In the case of a windmill, or other chiral devices, if time
is reversed the windmill `will not be able' to move in the opposite direction
because of the shape of its rotors, which is still pushing it in the
same direction.
What does move backward is the {\em mirror-image} of the same windmill.
Thus, in order to obtain a complete-reversal, it is also necessary `to reverse'
space.  In the case of an electric cell, by reversing the flow of time the
current `does flow' in the opposite direction even though the electrochemical
potential is still acting in the same direction, which contradicts the sense of the
current motion.  In order to reverse the flow of the current it is necessary
to {\it interchange} the  electrodes, i.e. to reverse also the sense of the electrochemical
potential.  To summarize on this symmetry consideration, CI is not a time-reversal
process but rather of {\em space-time} invariance, and this is so because of the
chiral nature of such an interaction.  Let T represent a time-reversal operation
and P is a space (or parity) reflection operation, then the main conclusion of this
argument is that CI does not obey T-invariance but rather PT-invariance.

Moving back to CI on a molecular level, it can readily be deduced that such a symmetry
consideration may have deep impact on the thermodynamical aspects of CI.
In the absence of T-invariance on a molecular scale it means that CI
does {\em not}
obey the classical detailed balance principles [24].  Detailed balance
implies
that at equilibrium the number of occurrences of each chemical reaction in the forward
direction is the same as that in the reverse direction.  In other words, equilibrium
does  exist not only macroscopically but also for each individual microscopic
reaction.  A necessary condition for detailed balance is the conservation of
time reversibility, which is not being obeyed by CI.  Moreover,
the persistence of
CI in one selected direction prevents the molecular system from reaching thermal
equilibrium.  This conclusion is related to the {\em ergodic assumption} introduced
by Boltzmann over a century ago. According to this assumption, any property taken as an
average over a large ensemble of particles within a closed box, being at thermal
equilibrium with its environment, can also be obtained for a single particle under
the same conditions, if averaged over a sufficiently long time.  The behaviour
of CI presents, therefore, a {\em non-ergodic} system.  The origin for this is
the generation of an intrinsic molecular perturbation that is moving
in one selective direction out of two possible ones,
as a result of a continuous interaction with random ionic motion
in the solvent at
the close vicinity of the molecule. Such a behaviour is uncommon for ordinary molecular
systems and it is the consequence of several structural details listed above,
in particular, its chiral structure as well as the presence of a microscopic
interface.

\noindent{\bf 4b.  Non-Ergodicity and Evolution}

The subject of thermal equilibrium and ergodicity may become of problematic nature
dealing with systems that contain more sophisticated elements.  As long as
relatively simple objects, such as mass points, or plain molecules are
concerned, the ergodic assumption seems to be fulfilled in general.  The problem arises
when less simple systems are involved, in particular, those which do not obey
time reversal invariance.  For example, suppose that an aqueous solution reaches thermal
equilibrium and then a bacterium is thrown into it.  A quasi-microscopic system,
such as a living bacterium, is certainly not time-reversible in its function,
and as long as it stays alive it also does not reach thermal equilibrium, i.e. it is
a {\em non-ergodic} system.  A similar conclusion can be reached for any
microscopic living system that happens to be immersed in a macroscopic surrounding
being at thermal equilibrium.  The question that can be asked now is:
``what is the
{\em smallest}  system that can remain at a non-ergodic state within a macroscopic
system being at thermal equilibrium".  Such a question cannot be readily
answered.  It seems reasonable to assume that any sufficiently complex microscopic
system, or molecule, that can perform in a `machine-like' mode of operation, being
time-irreversible in nature, will remain non-ergodic under such a
condition for
as long time as it does not break down or decompose into simpler and smaller
elements.  Soluble proteins, as well as other sufficiently complex
biomolecules,
seem to comply with this requirement.  It seems to be the case, and this is so
to the best knowledge of the author, that only `machine-like' biomolecules can be
considered as microscopic non-ergodic systems.  It may well be the case that also
 microelectronics systems may achieve such a state with further development.
In view of this it becomes of much interest to point out that non-ergodicity may
well be a phenomenon related to the microscopic level of function of life in
nature.

Another question of much interest concerns the possible significance of non-ergodicity
in nature.  Microscopic systems or molecules that reach easily thermal equilibrium
do not seem to be liable for any process of intrinsic organization. Another
significant aspect of the function and operation of complex molecular systems has
to do with its usefulness.  Any time reversal activity can be regarded as
completely ineffective from this view point of usefulness.  For this very reason the
breaking of time-reversal invariance of processes becomes necessary for a practical operation
of any useful value.  All machine operations are time-irreversible, and these
include microscopic `machines' as well, such as molecular
proteins.  Non-ergodicity becomes a feature closely related to time-irreversibility
for sophisticated microscopic systems.  Upon combining together
these various features, it
seems reasonable to deduce that the process of biomolecular evolution in nature
is closely related to them.  For this reason it becomes plausible to assume that
non-ergodicity is to be regarded a necessary condition for molecular evolution [11].
The CI hypothesis has similar features to those described hereby, and it is based on well
recognized structural details as well as on external source of random ionic motion,
which provides for a physical well explicable model.  For these reasons it is anticipated
that CI may become of major value for better understanding of the function of operative
biomolecules, soluble proteins in particular.  This is so, provided that more
experimental support is to be found for its viability.

\noindent{\bf 4c.  Order of Magnitude Estimate}

Another aspect of CI perturbation concerns the size of its energy content, being
of the order of ($10^{-1} - 10^{-2})$kT per chiral element, which is rather
small in comparison to thermal energy, being of the order of kT.  The significance
of such subthermal perturbation is not relevant to the size of its energy content,
but rather to its degree of ordering or sophistication.
The energy content of a biomolecule is
considerably larger, and this can serve as source of energy for its
operational functions.
The energy associated with CI can become operational upon `switching on' and
`switching off' the active groups in the molecule.  In other words, CI may be relevant
to the control mechanism of biomolecular function,
rather than to its operational function,
which requires much more energy.
A similar comparison of amounts of energy can be made between a robot and a
computer that controls its activity.  The energy needed for control is
considerably smaller than that required for the function of a robot, although its
degree of sophistication is rather impressive.  The relatively low energy content
of CI has an additional advantage as well.  Being of considerably lower energy
than typical thermal energy of the order of kT, does help to increase its length of
relaxation time $\tau_R$ which contributes to the persistence of CI.  The
reason for this effect is related to the presence of a large population of
energy modes of similar energy levels, as is the case of thermal energy, which
decreases their `life-times' lengths, owing to the high liability of interactivity
among such modes.

\noindent{\bf 4d. Possible Magnetic Effect}

This concerns the possibility of CI to induce magnetic fields along the axes
of coils of the $\alpha$-helix segments of which the globular protein consists.
The direction of this magnetic field follows the axis of each segment separately and
its magnitude is crudely estimated to be of the order of 0.01T, or 100 gauss.
It may well be too early to elaborate on the possible practical significance of such
an intrinsic molecular magnetic field.  However, it has already been emphasized
by Steiner and Ulrich [24] that magnetic fields can have significant effects on the
polarization of electronic and nuclear spins during chemical reactions, which may
considerably affect chemical yields and kinetics. It is important to emphasize that the
{\em size} of magnetic fields, applied externally to chemical reactions, has negligible
effect on these reactions.  It is also important to point out that in contrast
to regular chemical reactions, where an external magnetic field is applied
in arbitrary direction with respect to any molecule participating in such a
reaction, the intrinsic field induced by CI happens to be acting at the optimal
location and direction where it is needed.  This may well become a crucial feature of
CI, which can efficiently contribute to the autocatalytic function of
soluble proteins.  Another interesting feature of such an intrinsic magnetic field
induced by CI, is related to the description of `chiral favourable environment'
introduced by Barron [25], who proposes to apply a combination of electric
and magnetic fields parallel to one another in order to generate an enantiomeric
 excess in a chiral synthesis performed in their close presence.  Actually,
such an intrinsic combination exists in the globular state of soluble proteins
in the presence of CI.  The electric field is generated by the electric dipole moment
that exists along the peptide bond that comes into touch with the solvent
surrounding the globular protein, whereas the magnetic field is induced in
parallel to it by CI along the axes of the coils of the $\alpha$-helix.

\noindent{\bf 5. Experimental Verification and Support}

\noindent{\bf 5a. Background} \\
The presence of CI perturbation along one selected direction may produce an additional
effect.
Although CI is largely still a hypothesis awaiting for an experimental
verification,
there exist several pieces of evidence supporting its validity.  Let us first indicate
that its very existence is based largely on general physical principles which are
hard to refute, as well as on general symmetry arguments.  All macroscopic
examples are based on such considerations.  In addition, it is important to notice
that CI is a special and uncommon molecular phenomenon, which requires a set
of structural constraints for its possible validity, in analogy to macroscopic
chiral devices.  These include the presence of an electric dipole moment, as
well as an interface separating between the active medium in the solvent and
the inner part of the molecule.  All these happen to exist for soluble proteins.
Before looking at a possible and practical experiment, we note two specific difficulties
that may affect the observation of CI. The first one concerns the size of the effect,
which is quite small, and the second is the fact of its being an intrinsic molecular
event, which limits its experimental accessibility.  The second difficulty
concerns proteins rather than amino acids, which are of more open structure.  On
the other hand, there are good reasons why it is desirable to observe CI.  Such
an effect can provide for a better understanding of the complex nature and
operation of biomolecular function.

In the lack of any direct experimental evidence for the validity of CI it may be helpful to look
for existing experimental results that may bear certain close relation to this
phenomenon.  For this very purpose an {\em indirect} experiment has
been proposed [17], based on an assumption that CI is a necessary condition for
enzymatic activity in soluble proteins.  Such an assumption, reasonable
as it may sound, presents a certain constraint that at best can provide
for a strong positive experimental support of CI, rather than
a direct verification
of its validity.  Such an experiment was actually performed by Careri et al.
[26,27].  This experiment concerns the effect of dehydration on the protonic
motion throughout the hydration layers around soluble proteins.  The amount
of water surrounding each protein molecule is of crucial importance for free
protonic motion around this molecule. This is associated with the so-called
`percolation transition' involving water clusters around the protein molecule.
By dehydrating the water layers beyond the percolation transition, the
protonic motion stops, and so does also, simultaneously, the enzymatic activity
of the molecule.  As soon as there exists enough water within the hydration
layer surrounding the molecule, free protonic motion, or mobility, becomes
possible and resumes again.  This in turn causes also the onset of
enzymatic activity in the protein
molecule.  Protonic motion is identical in fact, with free ionic
mobility being a
necessary condition for CI in soluble proteins.  This fact
links closely between chiral interaction and these experimental results.
The close causal connection between protonic, or ionic, mobility and enzymatic activity
in soluble proteins, fits precisely with the assumption of the existence of CI.

\noindent{\bf 5b. Enantiomeric Excess Experiment}

A direct experiment to observe CI has already been proposed [11,17,28].  In
order to perform such an experiment, it is necessary to make use of some
observable property that is generated by this effect.  Such is the case with
amino acid molecules attached to the water-air interface.
For this purpose a
racemic solution of a hydrophobic amino acid, such as tryptophan,
may be employed,
as is shown schematically in Figure 6.  The side chain R attaches
itself to the
water surface, and CI occurs around the loop NH$^+_3$.COO$^-$.C, which
contains a zwitterion where an H-bond may exist between NH$^+_3$ and COO$^-$.
The ring of bonds may also contain a water molecule that relaxes the angular
strain of the bonds [28]. CI becomes possible there, close to the water-air
interface, owing to the gradient $\vec{\bigtriangledown} c\neq 0$ of the ionic
strength $c$ across the ring of bonds at the water surface.  CI induces there
a magnetic dipole moment $\vec{\mu}$, which has opposite components for the
L and D enantiomers, respectively, with regard to the normal to the water
surface.  Next, an external magnetic field $\vec{B}$ is applied normal to the\
water surface and this interacts with $\vec{\mu}$ with energy $E_\mu$:
\begin{eqnarray}
E_\mu = \pm \vec{\mu}.\vec{B}. 
\end{eqnarray}
where the $\pm$ signs depend on the L or D chirality respectively.  This energy
adds to the hydrophobic energy $E_h$, i.e. $E=E_h+E_\mu$ or:
\begin{eqnarray}
E=E_h\pm \vec{\mu} . \vec{B}, 
\end{eqnarray}
which results in an energy difference of $2E_\mu$ between the two enantiomers.
This, in turn, creates a small population change $\Delta n$ between the two
enantiomers according to Boltzmann law:
\begin{eqnarray}
\Delta n = n_L-n_D 
\end{eqnarray}
which depends on the direction of $\vec{B}$.  In order to estimate the size
of this effect, i.e. of $\Delta n/n$, where $n=n_L^0=n^0_D$ is the
racemic population, it is necessary to estimate the magnetic dipole moment
$\mu$, and this is given [17] by: $\mu \approx (10^{-1} - 10^{-2})\mu_B$, where
$\mu_B$ is the Bohr magneton.  Then
\begin{eqnarray}
\frac{\Delta n}{n} = \frac{2\mu B}{kT} \cong 10^{-3} \ \
({\rm for} \ \mu\cong10^{-1}\mu_B) 
\end{eqnarray}
and B is of the order of a few teslas.

The measurement itself can be performed by removing monolayers from the water
surface [29] that contains the amino acid population and then by measuring
their optical activity elsewhere, not in the presence of the magnetic field.
Other advanced technologies may also be available for this purpose [30].

The physical reasoning behind this experiment resembles the reasoning leading
to the natural selection of the L-anantiomer of amino acid in terrestrial
biomolecular evolution.  The mechanism may well be the same, but instead of
an applied magnetic field there exists the vertical component of the
terrestrial
field.  It is suggested that this process could have happened during the
prebiotic era over a localized region of the earth, where the vertical
component of the terrestrial magnetic field had a well defined component over
a sufficiently time length, so that one enantiomer, presumably L, had a slight
energy advantage over D.  The difference in energy is rather small, being of the
order of $(10^{-8} - 10^{-9})kT$. This energy estimate, although quite small, is
still considerably larger, by some 7 or 8 orders of magnitude, than the weak
nuclear current (WNC) interaction mechanism proposed by Kondepudi and Nelson
[31] for the natural selection of the L-anantiomer.  The present mechanism
does not give an {\em a priori} advantage to either L or D enantiomer.  This
advantage is an accidental outcome of the direction of the normal magnetic
field of earth that happened  to exist at the specific region of
the ocean where
and when natural selection happened to occur.  Recently, Barron [25] has proposed
to apply a `chiral favorable environment' of an electric and magnetic fields parallel
to one another, which may prefer one enantiomeric chirality over the other.
In order to perform such an experiment, Barron proposes also to  rotate
charges mechanically, which would make it similar, though considerably more
awkward, to the approach of CI. \\

\noindent{\bf 5c. Detection of Associated Magnetic Effects}

Apart from a possible enantiomeric excess effect, owing to an external
magnetic field applied on a racemic solution, there exist also other magnetic effects
of CI that can be detected experimentally.  These have already been indicated above
for the case of soluble globular proteins.  Such a magnetic structure
may be detected,
perhaps, by polarized neutron scattering.  Another quite fascinating
possibility
is related to the SQUID apparatus, an abbreviation of Superconducting Quantum Interference
Device [32], which can detect extremely small magnetic fields with resolution of
the order of one part in 10$^{11}$ of the earth's magnetic field, or of femtotesla
$(10^{-15}$ tesla).  This is so in addition to its possibility of detection very
minute changes, or differences, in magnetic fields.  For this purpose it is
suitable to prepare monomeric solution of a single enantiomer, say L, of a
hydrophobic amino acid in a sufficiently large concentration.  Such a solution will
create a magnetic monolayer at the water-air interface.  This is due to the
presence of a surface of magnetic moments induced by CI, all having parallel components
in the same direction normal to the water-air interface. (See figure 6). Such
a magnetic monolayer may generate a sufficiently strong magnetic field, normal
to this surface, to be detectable by the SQUID.  The size of the magnetic
field produced by, say, 10$^{12}$ molecules of amino acid per 1 cm$^2$ of this
surface,
\newpage
\begin{figure}[htb]
\centerline{\epsfxsize=10cm \epsffile[94 332 484 556]{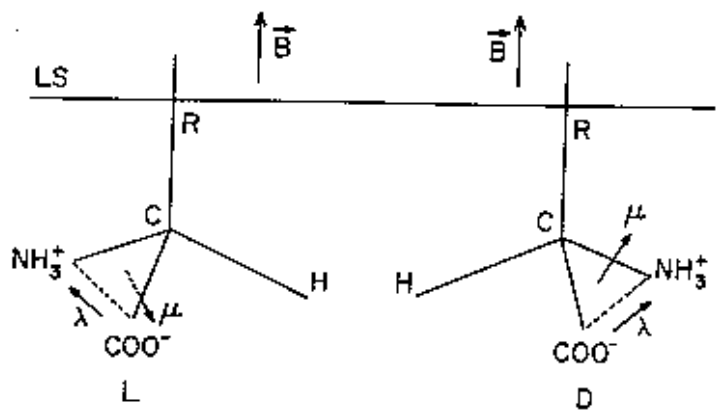}}
\caption{Two hydrophobic amino acid enantiomers L and
D are shown. The side chain R penetrates the water-air interface LS.
$\vec{B}$ is an external magnetic field applied normal to LS and interacts
with the induced magnetic moments $\vec{\mu}$. Due to the opposite
orientations of $\vec{\mu}$ with respect to $\vec{B}$ to L and D
respectively, there exists a small energy difference of $2 \vec{\mu}.
\vec{B}$ between the two enantiomers, at the water-air surface LS.}
\end{figure}
\newpage

is estimated to be of the order of $10^{-12}$ to $10^{-14}$ tesla
at a distance of 1 mm above the surface.  Such a magnetic field can,
hopefully, be detected by a SQUID [32]. \\

\noindent{\bf 6. The Reality of Chiral Interaction}

        The main motivation of the present treatment of CI comes form fundamental
questions such as: "why are the molecules of life chiral?" or, more specifically,
"does chirality offer any biological advantage to biomolecules?".  CI may provide
for a positive answer to such questions. For this very reason it has been helpful to
inspect macroscopic chiral devices and draw conclusions for chiral molecules by
pure analogy. Two modes of operation were found for CI, namely, massive rotational
motion of the chiral device and a circular flow of massless perturbation, such
as a current, throughout the device.  Of these, only the massless mode of
motion is found suitable for molecular CI.  Moreover, given certain structural details,
all happen to exist in soluble proteins, certain chiral molecules have the
capability of interacting with randomly moving ions in a solvent.  This interaction
produces an intrinsic perturbation that moves in one preferred direction out of
two possible ones.  Such a capability does not exist for achiral molecules.
This mode of perturbation within a  chiral molecule gives rise to a certain degree
of non-ergodicity, which may also be regarded as a certain amount of negative
entropy.  It is tempting to compare this to the somewhat naive conclusion of
Shrodinger [33] in his book {\em What is Life?}, where he claims for such a
phenomenon that: `it feeds on negative entropy'.  In the present context CI
enables proteins to reduce their entropy, i.e. to avoid thermal equilibrium by
a minute amount, which is to be regarded as a mode of `subthermal organization'.
It is hereby conjectured [11,17] that  such a phenomenon of non-ergodicity is
to be considered as a necessary condition for evolutionary process
in biomolecules.
It is also important to observe that such features require CI to
become a microscopically
time-irreversible [34] phenomenon.  In fact, if proteins are to be
regarded as `biological machines', there must exist an element of
time irreversibility in their function.

These arguments are rather general and still somewhat speculative, since as
yet CI has not been proven to be real on molecular level.  Moreover, it may be
quite plausible that such arguments are crucial to the function of proteins, though
it is still not easy to describe exactly how.  We still need a much better
insight into their function.  However, it is  important to indicate that if
CI is indeed of significance for protein function, then it can be useful to ascribe
biological significance to various structural features of soluble proteins that
otherwise would remain plain facts.  Among these features the chirality of the
protein molecule can be mentioned,
in particular the charge distribution on the peptide H-bonds
connecting the loops along the $\alpha$-helix coils, which are of physical chiral
nature.  In addition, this also offers a certain meaning to the presence of
nitrogen atoms in proteins and amino acid constitution, where the asymmetry between N
and O atoms is responsible for the dipolar charge on the peptide H-bonds.
By comparison, N atoms are totally absent in hydrocarbons.  Also structurally
meaningful is the existence of closed loops of chemical bonds in proteins,
which provide for a means of delivering the CI perturbation around the
molecule.  This holds for $\beta$ sheets as well [19].  An additional
meaningful
fact is the parallelism, or consistency, of all dipolar moments
along the $\alpha$-helix, which
adds up to a total dipole moment in each separate segment of a globular
protein along the narrow zone that happens to be in contact with the solvent.
(see figure 4).  All these structural details have already been considered in the
assumptions leading to the proposal of CI, so that their biological
significance is not surprising.  More surprising is the fact that soluble proteins
must become globular before they can function as enzymes.  This fact has not been
considered previously.  It links up neatly with the necessity of an interface
for the operation of CI.  Another fact of biological importance is the presence
of free mobile ions in a solvent, which leads to the preference of
physiological, rather than pure or distilled water, as the solvent material.
These mobile ions are necessary
for generating CI in biomolecules.  Most relevant and significant, so far,
is the contribution made by the experiment [26,27] of Careri et al., which reveals
the connection between protonic motion in the hydration layer of proteins and the
onset of enzymatic activity.  This experiment provides for an impressive support for the
CI hypothesis.

Apart from these facts there exist two related points that are of biomolecular
evolutionary significance.  One of these has to do with the possibility that
CI was playing an important role in the natural selection of the L-anantiomer
of amino acids. The other is the evolutionary significance of the non-ergodic nature
of CI.  Although in nature there hardly exists any thermal equilibrium, it seems
very likely that non-ergodic systems, such as proteins, become much more susceptible
to evolutionary processes in comparison to systems that readily reach
thermal equilibrium.  it is also interesting to indicate that any living
system is non-ergodic as well.

Another aspect of CI concerns the amount of energy involved in such a process.
This energy is rather small, subthermal in effect, which may evoke criticism
as to its significance. Such criticism is rather common among scientists
who attribute significance to energy according to its size. What is
significant in complex systems such as proteins, may well be the quality,
or the degree of sophistication of energy, rather than its size.
For instance, information
transferred by electromagnetic wave involves certain modulations of this
wave and their respective energy is much less than the energy of the wave
itself, though its content is of major importance and usefulness. Another
example is the small amount of energy required to switch on and off a much
larger source of energy. This is to be regarded as a mode of control energy
which may be of main significance in CI.  In the case of a robot controlled
by a computer, we have a similar example. The quantity of energy required
by a robot is considerably larger than that of the computer, but without
this small quantity of energy, the robot cannot function coherently. Another,
rather cruel example, concerns the size of energy change that occurs over
the small interval during which a creature dies. This change in energy is
very small, but its significance is enormous. This example may well indicate
the significance of the content of such a small amount of energy. It is
interesting, even fascinating, to point out that in the case of CI all this
ordered energy comes from chaotic random motion of ions, and this is mainly
due to the chiral nature of CI.  In addition to all these, it may be
relevant to indicate that the density of energy states for subthermal energies
in molecular systems is considerably smaller than that within the thermal
range, which makes subthermal energies less convertible, or dissipative,
than thermal modes of energy, and therefore functionally more efficient.

Another significant feature of CI is its time irreversibility. It is
important to emphasize that any time-reversible operation is meaningless
when regarded from aspects of usefulness. Any productive machine function
becomes automatically time-irreversible if its operation is of any merit.
For instance, even information transfer, which usually requires very little
energy, is totally a time-irreversible process. The source of time
reversibility in physics comes from the nature of the Newtonian mass point
which is an ideally symmetric object. If instead of such a symmetric point
mass, an elementary physical chiral object [11] is to be employed, then
it becomes likely that instead of time-reversal invariance the rule of
PT-invariance will dominate. This may well be the main source, or reason,
for the biological advantage of chirality.

In conclusion, let it be reminded that in spite of strong but indirect
support, such as that of the experiment of Careri et al. [26,27], the CI
hypothesis is still in its infancy and requires much more insight,
understanding and development, not to mention additional support from
direct experiments as discussed here and elsewhere [11,17,28].
\pagebreak

\noindent {\bf 7. Conclusions} \\

The phenomenon of chiral interaction has been described and treated in detail
in this article. Various uncommon features of such an interaction are
described and discussed. It is also claimed that chiral interaction may
well be of significant biological advantage, and this is due to its possible
linkage and relevance to enzymatic activity of soluble proteins. Another
reason for such an advantage is that chiral interaction may be the source
of non-ergodicity in biological molecules, which might be relevant to the
process of biomolecular evolution in nature.
\vspace{0.5cm}

\noindent {\bf Acknowledgement} \\

The author wishes to thank Dr. Robert M. Clegg for his help in discovering
the work of Careri et al. He also wishes to thank Ms. Elizabeth Youdim,
and Ms. Gila Etzion for their considerable help in completing this article.
\pagebreak

\noindent {\bf References:}

\begin{enumerate}
\item F. Arago; ``Memoires de la Classe des Sciences Math. et Phys. de
l'Institut Imperial de France'', Part 1, p. 93, (1811).
\item J.B. Biot; ``Memoires de la Classe des Science Math. et Phys. de
l'Institut Imperial de France'', Part 1, 1, (1812).
\item L. Pasteur; Ann. Chim. 24, 457 (1848).
\item W.T. Kelvin; ``Baltimore Lectures'' (C.J. Clay \& Sons, London, 1904).
\item P.G. Mezey; J. Math. Chem. 17, 185 (1995),
Comput. Math. 34, 105 (1997), and references therein.
\item G. Gilat and Y. Gordon; J. Math. Chem. 16, 37 (1994).
\item G. Gilat; J. Phys. A, 22, L545 (1989), {\it ibid.} Found. Phys. Lett.
3, 189 (1990).
\item A.B. Buda and K. Mislow; J. Am. Chem. Soc. 114, 6006 (1992).
\item H. Zabrodsky and D. Avnir, J. Am. Chem. Soc. 117, 462 (1995).
\item A.B. Buda, T.P.E. Auf der Heyde and K. Mislow; Angew. Chem. Int.
Ed. Engl. 31, 989 (1992).
\item G. Gilat; The Concept of Structural Chirality, in ``Concepts in
Chemistry'' Ed. D.H. Rouvray (Research Studies Press and Wiley \& Sons,
London, New York, 1996), p. 325.
\item G. Gilat; J. Math. Chem. 15, 197 (1994).
\item G. Gilat; Chem. Phys. Lett. 121, 9 (1985).
\item G. Gilat and L.S. Schulman; Chem. Phys. Lett. 121, 13 (1985).
\item G. Gilat; Chem. Phys. Lett. 125, 129 (1986).
\item G. Gilat; Chem. Phys. Lett. 137, 492 (1987).
\item G. Gilat; Mol. Eng. 1, 161 (1991).
\item A.S. Davydov; ``Biology and Quantum Mechanics'', (Pergamon Press,
Oxford, 1982).
\item For more details concerning the steric structure of globular proteins,
see R. Huber and W.S. Bennett Jr. in ``Biophysics'' Eds. W. Hoppe,
W. Lohmann, H. Markl \& H. Ziegler, (Springer-Verlag, Berlin, 1983),
p. 372. Globular proteins contain also $\beta$-sheets which are capable
too of generating CI in solvents.
\item H. Tschesche; in ``Biophysics'', Eds. W. Hoppe, W. Lohmann, H. Markl
\& H. Ziegler, (Springer-Verlag, Berlin, 1983), p. 37.
\item H. Haken; ``Synergetics - an Introduction'', (Springer-Verlag, Berlin,
1977).
\item L.D. Barron; ``Molecular Light Scattering and Optical Activity'',
(Cambridge University Press, Cambridge, 1982).
\item L.E. Reichl; ``A Modern Course in Statistical Mechanics'', (University
of Texas Press, Austin, TX 91980, 1980), p. 468.
\item U.E. Steiner and T. Ulrich; Chem. Rev. 89, 51 (1989).
\item L.D. Barron; Science 266, 1491 (1994), and references therein.
\item G. Careri, A. Giasanti and J.A. Rupley; Phys. Rev. A37, 2703 (1988).
\item J.A. Rupley and G. Careri; Adv. Protein Chem. 41, 37, (1991).
\item G. Gilat; Chem. Phys. 140, 195 (1990).
\item M. Pomeranz, F.H. Docol and A. Segmuller; Phys. Rev. Lett. 40, 2467
(1978).
\item G.L. Gaines, Jr.; ``Insoluble Monolayers at Liquid-Gas Interfaces'',
(Wiley \& Sons, New York 1966), p. 125 and references therein.
\item D.K. Kondepudi and G.W. Nelson; Phys. Rev. Lett. 50, 1023 (1983),
{\it ibid.} Phys. Lett. 106A, 203 (1984).
\item J. Clarke, Sci. Am. 271, 46 (1994).
\item E. Schrodinger; ``What is Life'' (Cambridge University Press,
Cambridge, 1944),
\item P.C.W. Davies; ``The Physics of Time Asymmetry'', (Surrey University
Press, 1974).
\end{enumerate}
\vspace{0.5cm}


\end{document}